\documentclass[%
 reprint,
 amsmath,amssymb,
 aps
,longbibliography]{revtex4-1}

\usepackage{braket}
\usepackage{graphicx}% Include figure files
\usepackage{dcolumn}% Align table columns on decimal point
\usepackage{bm}% bold math
\usepackage{xcolor}

\begin{document}
--
%\preprint{APS/123-QED}

\title{Enantiosensitive exceptional points}
% Force line breaks with \\

\author{Nicola Mayer$^{1}$,
Nimrod Moiseyev$^{2}$,
Olga Smirnova$^{1,3}$}

\affiliation{
$^1$Max-Born-Institute, Max-Born Strasse 2A, 12489 Berlin, Germany \\
$^2$Schulich Faculty of Chemistry and Faculty of Physics, Technion, Haifa, 32000, Israel\\
$^3$ Department of Physics, Technical University Berlin, 10623 Berlin, Germany}

\date{\today}

\begin{abstract}
We show that the position of the exceptional points (EPs) in the parameter space of a chiral molecule coupled to the photoionization continuum by a three-color field is enantiosensitive. Using a minimal model of a three-level system driven by a three-color field to form a cyclic loop transition, we investigate the enantiosensitivity of the EPs with respect to the system parameters and exploit the asymmetric switch mechanism associated with the encirclement of an EP in parameter space in an enantio-selective way. Our work paves the way for future applications of enantiosensitive EPs in chiral systems.
\end{abstract}

\maketitle

Exceptional points (EPs) are points in the parameter space of non-Hermitian Hamiltonians where at least two eigenvalues and corresponding eigenvectors coalesce \cite{Heiss:2012aa}. Their remarkable properties derive from their topological structure of a branching point connecting the Riemann sheets of the coalescing adiabatic states, explored experimentally in \cite{PhysRevLett.86.787,Gao:2015aa}. For example, encircling an EP in parameter space leads to a switch between the eigenstates and eigenvalues of the adiabatic solutions, the so-called \textit{adiabatic flip} effect \cite{Heiss:1999,KAPRALOVAZDANSKA2022168939}. Due to the inevitable non-adiabatic transitions in the dynamical evolution around the EP, this leads to an asymmetric behaviour where the final state depends only on the sense of encirclement of the EP, known as asymmetric switch mechanism (ASM), explored in the atomic and molecular case in Refs. \cite{Uzdin:2011aa,Gilary:2012,Zdanska:2014aa,PhysRevA.88.010102} and in other systems such as waveguides in Refs. \cite{Doppler:2016aa,Xu:2016aa}.\par
EPs are also critical points connecting $\mathcal{PT}$-symmetric and $\mathcal{PT}$-broken regions of a non-Hermitian system \cite{Ozdemir:2019aa,Moiseyev2018}.
The connection of the concept of EPs with chiral, i.e. $\mathcal{P}$-breaking, systems seems therefore a particularly promising and interesting one, yet it has been hardly and only recently explored. For example Ref. \cite{PhysRevB.101.214109} has studied the influence of PT-symmetric chiral metamaterials on the propagation of circularly polarized light; Ref. \cite{PhysRevLett.124.083901} has shown that EPs can be exploited to create superchiral fields with definite handedness from photonic crystal slabs, enhancing optical activity in chiral molecules; similarly, Ref. \cite{PhysRevA.105.053711} has proposed using a high-Q cavity tuned to an EP for enhanced enantiomeric discrimination.\par
While the application of these works for sensing methods of molecular chirality is very promising, here we take a different and novel approach, showing that position of the EPs in parameter space is sensitive to the handedness of the non-Hermitian chiral system they stem from.
We do so by adopting a minimal open three-level model, meant to represent a chiral molecule coupled to the photoionization continuum via a three-color laser field. We demonstrate that for an appropriate choice of laser parameters one can tune EPs of opposite enantiomers toward different positions in parameter space. This result allows us then to exploit EP-related effects for only one of the two enantiomers, such as the asymmetric switch. The generality of this new concept opens a way for completely novel schemes of enantio-separation.

%\section{Model chiral non-Hermitian Hamiltonian}
%
\begin{figure}
\begin{center}
\includegraphics[width=6cm, keepaspectratio=true]{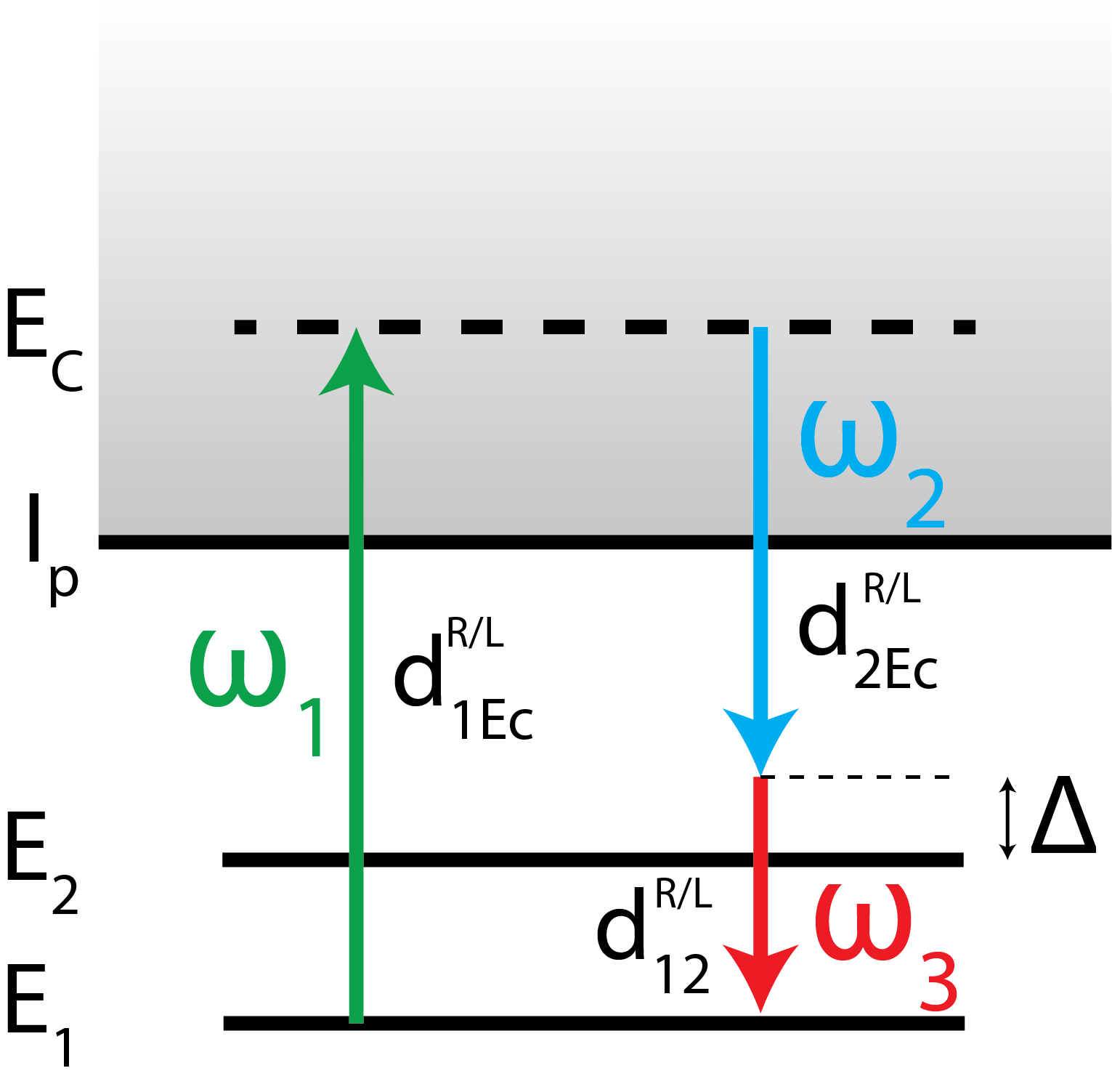}
\caption{The open three-level model representing a chiral molecule coupled to the photoionization continuum by a three-color laser field such that $\omega_1=\omega_2+\omega_3$. Opposite enantiomers are related by an inversion $\mathbf{d}^{R}=-\mathbf{d}^{L}$.}
\label{Fig0}
\end{center}
\end{figure}
We model a chiral molecule coupled to the photoionization continuum by a three-color laser field using a three-level system composed of two bound states $|1\rangle$ and $|2\rangle$ and a \textit{flat} continuum state $|E_C\rangle$ that does not include resonances in the vicinity of the transitions. In order to probe the chirality of such a three-level system, we need to form a closed loop of transitions using a three-color laser field, as shown in Ref. \cite{PhysRevLett.87.183002} for a bound-only system. Hence, we choose a three-color laser field with frequencies $\omega_1$ and $\omega_2$, which couple respectively the $|1\rangle$ and $|2\rangle$ bound states to the continuum $|E_C\rangle$, and an additional frequency $\omega_3$ which couples the two bound states via a one-photon transition. Choosing the laser frequencies such that $\omega_1=\omega_2+\omega_3$, we obtain a closed loop of transitions as shown in Fig. \ref{Fig0}. Note that if the system was achiral, i.e. $\mathcal{P}-$symmetric, then the loop could not be closed, as the one- and two-photon transitions would lead to final states with opposite parity. The handedness of the model system can be changed by inverting the dipole moments, i.e. $\mathbf{d}^{R}=-\mathbf{d}^L$.\par
The Hamiltonian of the driven system is $H(t)=H_0-\mathbf{d}\cdot\mathbf{E}(t)$, where $H_0$ is the field-free Hamiltonian with diagonal entries corresponding to the field-free energies of the states, $\mathbf{d}$ is the dipole operator and $\mathbf{E}(t)=\sum_i\mathbf{F}_i(t)\cos(\omega_i t)$ is the three-color field, where $\mathbf{F}_i=F_i(t)\mathbf{e}_i$ ($F_i(t)$ is the envelope, $\mathbf{e}_i$ is the polarization vector). As shown in the Supplementary Information \cite{SuppNotes}, inserting the ansatz $|\Psi(t)\rangle=\sum_ic_i(t)|i\rangle+\int dE_Cc_{E_C}(t)|E_C\rangle$, applying the Rotating Wave Approximation and adiabatic elimination of the continuum \cite{Fedorov:book,KNIGHT19901,Mayer:2020aa}, we obtain a reduced effective non-Hermitian Hamiltonian for the evolution of the two bound amplitudes, cast in matrix form as
\begin{equation}H_{R/L}=\begin{bmatrix}-\frac{\text{i}}{2}\Gamma_1 & V^{R/L}_{12} \\ V^{R/L}_{21} & \Delta-\frac{\text{i}}{2}\Gamma_2\end{bmatrix}\end{equation}
where we ignore for simplicity permanent dipoles in the bound states.
Here $\Gamma_i=\pi|\mathbf{d}_{i,E_c}\cdot\mathbf{F}_i|^2$ are the decay rates toward the continuum of the bound states, where $\mathbf{d}_{i,E_c}=\langle i|\mathbf{d}|E_c\rangle$ is the bound-free dipole matrix element. $\Delta=E_2-E_1-\omega_3$ is the detuning of the one-photon transition coupling directly the two bound states, and the off-diagonal coupling is explicitly given by
\begin{equation}V^{R/L}_{12}=-\frac{\text{i}}{2}\pi\left(\mathbf{d}_{1,E_C}\cdot\mathbf{F}_1\right) (\mathbf{d}_{2,E_C}\cdot\mathbf{F}_2)^*+\Omega^{R/L}_{12}\end{equation}
\begin{equation}V^{R/L}_{21}=-\frac{\text{i}}{2}\pi (\mathbf{d}_{1,E_C}\cdot\mathbf{F}_1)^*(\mathbf{d}_{2,E_C}\cdot\mathbf{F}_2)+\Omega^{R/L}_{21}.\end{equation}
Here the complex term represents the two-photon Raman-like coupling of the two bound states through the continuum (see Fig. \ref{Fig0}), while $\Omega_{12}^{R/L}=-\left(\mathbf{d}_{12}\cdot\mathbf{F}_3\right)/2$ is the Rabi frequency of the one-photon coupling. Crucially, a change in handedness $\mathbf{d}^R=-\mathbf{d}^L$ leaves the two-photon coupling unchanged, while the Rabi frequency changes sign. Accordingly, we label the Rabi frequency using the letters R/L (right/left). Eq. (1) is therefore an example of a chiral non-Hermitian Hamiltonian.\par
EPs are found when the eigenenergies $\gamma_{\pm}$ of the Hamiltonian (1) are degenerate, where $H_{R/L}|\phi^{R/L}_{\pm}\rangle=\gamma^{R/L}_{\pm}|\phi^{R/L}_{\pm}\rangle$. These are explicitly given by
\begin{equation}\gamma^{R/L}_{\pm}=\frac{\Delta-\text{i}(\Gamma_1+\Gamma_2)\pm\sqrt{\delta^{R/L}}}{2}\end{equation}
where
\begin{eqnarray}\delta^{R/L}&=&\Delta^2-\left(\frac{\Gamma_1+\Gamma_2}{2}\right)^2+4(\Omega^{R/L}_{12})^2\nonumber\\
&&-\text{i}\Delta(\Gamma_2-\Gamma_1)-4\text{i}\Omega^{R/L}_{123}\end{eqnarray}
Here we have defined the three-photon cyclic matrix element $\Omega^{R/L}_{123}=(\mathbf{d}^{R/L}_{1,\epsilon}\cdot\mathbf{F}_1)^*(\mathbf{d}^{R/L}_{2,\epsilon}\cdot\mathbf{F}_2)(\mathbf{d}^{R/L}_{1,2}\cdot\mathbf{F}_3)$, which encodes the handedness of the system since $\Omega^{R}_{123}=-\Omega^{L}_{123}$ (see Ref. \cite{PhysRevLett.87.183002} for the bound states only case). EPs are given by the condition $\delta^{R/L}=0$, and clearly, due to the presence of the $\Omega^{R/L}_{123}$ term, their position in the parameter space of the Hamiltonian depends on the handedness of the molecule. This is the core result of this paper. Obviously, if the loop is not closed ($F_i=0$ for any of the three fields), enantiosensitivity is lost because $\Omega^{R/L}_{123}=0$.\par
More explicitly, let us study the position of the EPs in the $(\Delta,\Omega_{12})$ parameter space. For each enantiomer, we find two EPs. Their position for the enantiomers is given by
\begin{eqnarray}\Omega^{EP_R}_{12}&=&-\Omega^{EP_L}_{12}=\pm\frac{\Gamma_2-\Gamma_1}{4}\\
\Delta^{EP_R}&=&\Delta^{EP_L}=\mp\sqrt{\Gamma_1\Gamma_2},\end{eqnarray}
where we see that for the left enantiomer the EPs are reflected on the $\Omega_{12}$ axis.
\begin{figure}
\begin{center}
\includegraphics[width=7cm, keepaspectratio=true]{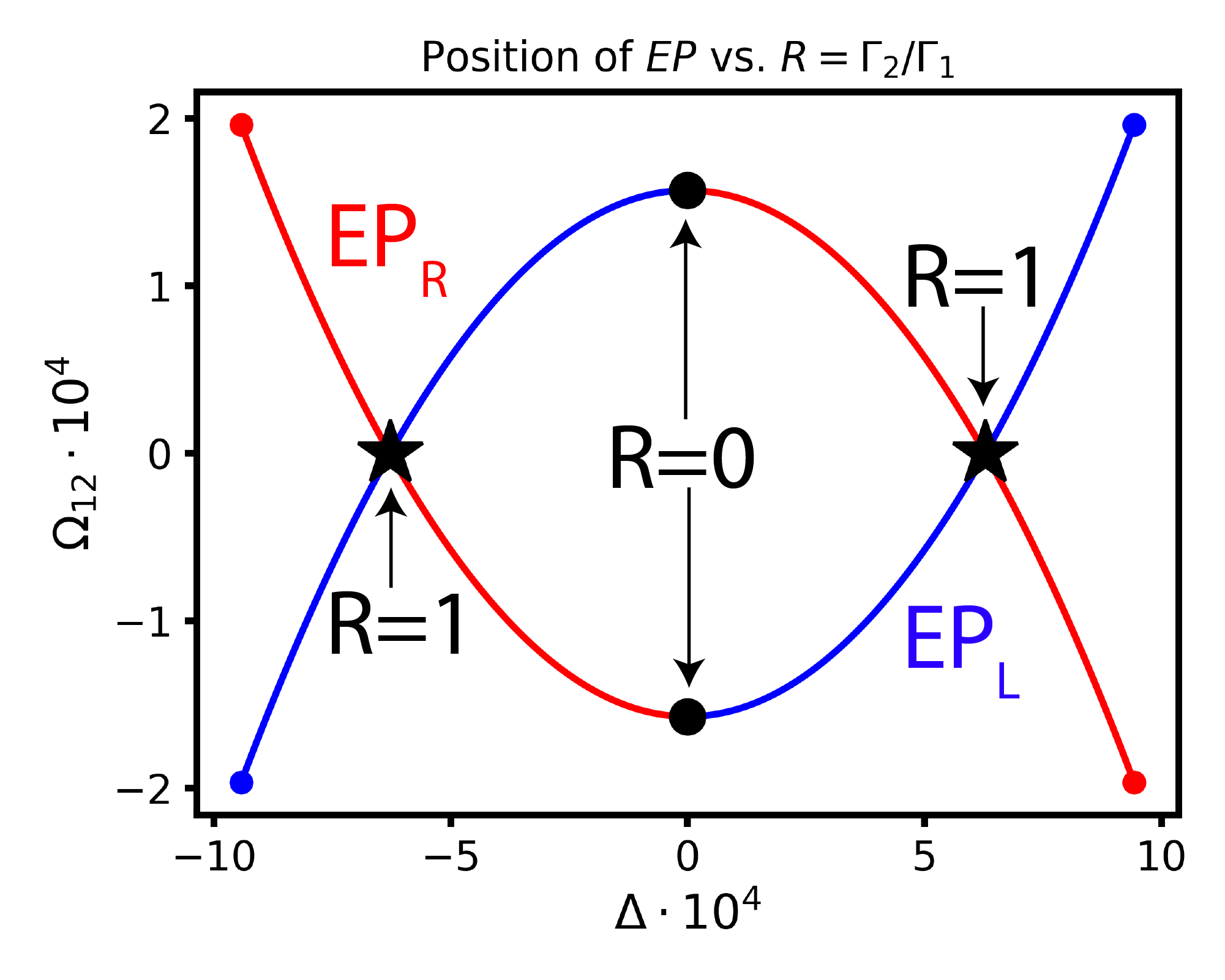}
\caption{Position of the EPs of the two enantiomers in the parameter space $(\Delta,\Omega_{12})$ for varying ratios $R=\Gamma_2/\Gamma_1$ when $\Gamma_1=6.2\cdot10^{-3}$ a.u.; the solid red and blue lines correspond respectively to the EPs of the right and left enantiomer. The black dots correspond to the case $R=0$, the black stars to $R=1$ and the colored dots to $R=2.25$.}
\label{Fig1}
\end{center}
\end{figure}
From Eqs. (6) and (7) we also can see that if the decay rates are equal $\Gamma_1=\Gamma_2$ the enantiosensitivity is lost and the EPs of the two enantiomer collapse to the same achiral position $\Omega^{EP}_{12}=0$. This is shown in Fig. 2, where the position of the EPs of the two enantiomers is studied in the $(\Delta,\Omega_{12})$ parameter space varying the ratio $R=\Gamma_2/\Gamma_1$ for fixed $\Gamma_1=6.2\cdot10^{-3}$ a.u.. At $R=0$ the loop is not closed ($\Gamma_2=0$) and the EPs of both enantiomers lie in the same position along the $\Delta=0$ axis. As soon as $R>0$, the EPs of the two enantiomers split from their initial position and enantiosensitivity is obtained. At $R=1$ ($\Gamma_2=\Gamma_1$) enantiosensitivity is lost again (see Eq. (6)), with now the EPs lying along the $\Omega_{12}=0$ axis. For any $R>1$, the EPs of the two enantiomers are separable in parameter space. This shows that we can optimally tune the field parameters in order to achieve as much separation as needed between the EPs of opposite enantiomers. Obviously, these results are valid for any two sets of parameters we choose to vary. For example in the Supplementary Information we show the EPs enantiosensitivity in the $(\Gamma_1,\Gamma_2)$ parameter space.

%\section{Enantiosensitive asymmetric switch}
%
\begin{figure*}
\begin{center}
\includegraphics[width=18cm, keepaspectratio=true]{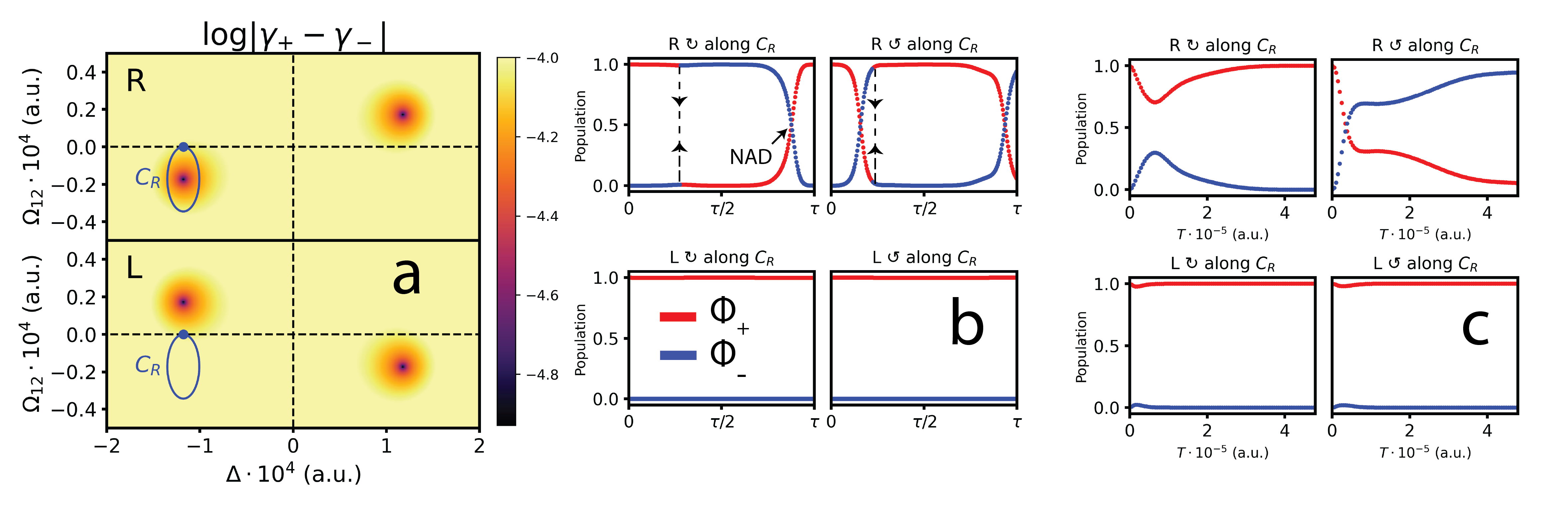}
\caption{Asymmetric switch mechanism in chiral systems. \textbf{a)}: Logarithm of the absolute value of the difference between the eigenenergies $\gamma_{\pm}$ for the right (top) and left (bottom) enantiomers in the $(\Delta,\Omega_{12})$ parameter space when $\Gamma_1=1.5\cdot10^{-4}$ a.u. and $\Gamma_2=8.8\cdot10^{-5}$ a.u.. The dark spots indicate the EPs, while the solid blue line shows the path $C_R$ that encloses an EP of the right enantiomer. \textbf{b)}: Dynamical evolution of the adiabatic states along the loop $C_R$ as a function of the scaled time $\tau=t/T$, when $T=4.78\cdot10^{5}$ a.u. and the population is initially in the $\phi_+$ adiabatic state; the solid red and blue lines indicate respectively the $\phi_+$ and $\phi_-$ adiabatic states. The dashed arrows indicate the time at which the system passes from one energy surface to the other, while NAD indicates the non-adiabatic transition. \textbf{c)}: Final population in the adiabatic states as a function of the loop time up to $T=4.78\cdot10^{5}$ a.u.; both in figures b) and c) the top plots shows the results for the right enantiomer, while the bottom plots show the results for the left enantiomer. Note that all populations are normalized to unity in order to aid visualization of the results.}
\label{Fig2}
\end{center}
\end{figure*}

We now exploit the enantiosensitivity of the position of EPs in the parameter space of a chiral molecule by encircling them in the parameter space and studying the resulting dynamics. When an EP is encircled, the quasi-energies and corresponding eigenstates of the Hamiltonian at the end of the evolution $t=T_{loop}$ have switched up to a phase, i.e. $\gamma_{\pm}(t=T_{loop})=\gamma_{\mp}(t=0)$ and $|\phi_{\pm}(t=T_{loop})\rangle=|\phi_{\mp}(t=0)\rangle\exp(i\Theta)$. Yet, due to the non-Hermiticity of the dynamics, the rate of the non-adiabatic transitions between the two adiabatic solutions has an exponential dependence on the imaginary components of the adiabatic energies, leading to unavoidable non-adiabatic transitions between the adiabatic states. Crucially, the sign of this transition rate depends on the sense of encirclement of the EP in parameter space \cite{Uzdin:2011aa,Gilary:2012,PhysRevA.88.010102}, and as a result, the ASM effect is obtained, where encirclements of the EPs lead to a final adiabatic state that depends only on the sense of encirclement, rather than on the initial conditions in which the system is prepared. In our case it is clear therefore that by an appropriate choice of the path in parameter space we can encircle the EP of a given enantiomer and selectively activate the ASM mechanism for one molecular enantiomer, while its mirror twin follows a completely different temporal evolution, as its EP is in a different position of the parameter space.\par
We verify this proposal by encircling the EP in the $(\Delta,\Omega_{12})$ parameter space, keeping the $\Gamma_i$ decay rates fixed. This corresponds to introducing a chirp into the laser field that couples the two bound states and varying its intensity, while also introducing a chirp to the laser field that couples state $|2\rangle$ to the continuum in order to keep $\Gamma_2$ fixed. The laser field that couples state $1\rangle$ to the continuum is kept fixed in both intensity and frequency. We use the decay rates $\Gamma_1=1.5\cdot10^{-4}$ a.u. and $\Gamma_2=8.8\cdot10^{-5}$ a.u. and choose the path labelled $C_R$ parametrized as
\begin{eqnarray}\Delta(t)&=&\Delta_{EP}+\rho\sin\left(2\pi\frac{t}{T}\right)\nonumber\\
\Omega_{12}(t)&=&\Omega^{EP}_{12}+\rho\cos\left(2\pi\frac{t}{T}\right)\end{eqnarray}

where $\Delta_{EP}=-1.718\cdot10^{-5}$ a.u., $\Omega_{EP}=-1.117\cdot10^{-4}$ a.u. and $T$ is the loop time. We choose the radius of the path to be $\rho=-\Omega_{12}^{EP}$, in order to start and finish the dynamics with absent direct one-photon coupling between the two bound states. In order to solve the TDSE, we first perform the transformation $a_i(t)=c_i(t)\exp\left(-\text{i}\int_0^{t}dt'\left(\text{i}\Gamma(t')/2+\Delta(t')/2\right)\right)$, where $\Gamma=\left(\Gamma_1+\Gamma_2\right)/2$, and obtain the Hamiltonian
\begin{equation}\mathbf{M}(t)=\begin{bmatrix}-\frac{\Delta}{2}-\text{i}\frac{\gamma}{2} & V_{12} \\ V_{21} & \frac{\Delta}{2}+\text{i}\frac{\gamma}{2}\end{bmatrix}\end{equation}
where $\gamma=(\Gamma_1-\Gamma_2)/2$. For the eigenstates basis, we choose the parallel transport basis (see Ref. \cite{PhysRevA.92.052124})
\begin{eqnarray}\mathbf{\phi}^r_{+}=\mathbf{\phi}^{l}_{+}=\begin{bmatrix} \cos(\theta/2) \\ \sin(\theta/2)\end{bmatrix}\nonumber\\
\mathbf{\phi}^r_{-}=\mathbf{\phi}^{l}_-=\begin{bmatrix} -\sin(\theta/2) \\ \cos(\theta/2)\end{bmatrix}\end{eqnarray}
where $\tan(\theta)=-2V_{12}/\left(\Delta+\text{i}\gamma\right)$. Here we have used the lowercase r/l to indicate right and left eigenvectors of the non-Hermitian Hamiltonian; note that the eigenvectors respect the c-product $\left(\phi^{r/l}_{-}|\phi^{r/l}_{+}\right)=0$. At the EP the two eigenvectors coalesce and self-orthogonality occurs $\left(\phi_{\pm}|\phi_{\pm}\right)=0$ \cite{moiseyev_2011}; in our case this is not relevant as we stay always sufficiently far away from the EP. We then solve the TDSE in the instantaneous basis for $t\in\left[0,T\right]$, and find the corresponding amplitudes of the adiabatic states by projecting the instantaneous solution on the adiabatic basis of Eq. (10). For the simulations, we set the initial conditions such that the initially the population is in the $\phi_+$ adiabatic state $a_+(0)=0$ (results for other initial conditions are reported in the Supplementary Informations \cite{SuppNotes}).\par
The results are shown in Fig. \ref{Fig2}. Fig. \ref{Fig2}a) shows the logarithm of the difference between the quasi-energies $\gamma_{\pm}$ in the $(\Delta,\Omega_{12})$ parameter space for the two enantiomers. The EPs of the two enantiomers are related by a mirror reflection along the $\Omega_{12}$ axis, as expected. The blue line indicates the $C_R$ path of Eq. (8). Results for paths enclosing an EP of the left enantiomer are shown in the Supplementary Information \cite{SuppNotes}. Fig. 3b) shows the dynamical evolution of the adiabatic states obtained after projection of the instantaneous solution for a loop time of $T=4.78\cdot10^{5}$ a.u., plotted as a function of the scaled time $\tau=t/T$. Note that the total population in the adiabatic states is normalized to unity in order to better visualize the results. The top plots in Fig. 3b) show the results for the right enantiomer, while the bottom plots correspond to the left enantiomer. For the right enantiomer, a clockwise encirclement of the EP leads to the final population being in the $\phi_+$ state, while a counter-clockwise encirclement leads to the final population being in the $\phi_-$ state. This result is independent of the initial conditions (see Supplementary Information \cite{SuppNotes}) and corresponds to the ASM effect. The dashed arrows at $\simeq\tau/4$ indicate the instant at which the system crosses the branch cut between the two adiabatic solutions, leading to a relabelling of the states. The acronym NAD indicates a non-adiabatic transition between the two adiabatic solutions. These are not indicated for the counter-clockwise encirlement for simplicity. The dynamical evolution in the left enantiomer shows instead that the final population is in the $\phi_+$ state independently from the sense of encirclement, since for this enantiomer no EP is enclosed. The results thus confirm that the ASM effect can be activated in an enantioselective way for same field parameters.\par
In order to study the regime of validity of the enantiosensitive ASM effect, we record the final (normalized to unity) population in the adiabatic states for varying loop time up to $T=4.78\cdot10^{5}$ a.u.; the results are shown in Fig. \ref{Fig2}c), which shows that ASM is seen already for $T\geq10^{5}$, with better separation of the final populations for increasing loop times. For the largest loop time, the final non-normalized populations in the right and left enantiomers are respectively on the order of $10^{-8}$ and $10^{-4}$ respectively. In an experiment, it would be ideal to maximize the residual population by an appropriate choice of the path in the parameter space, here beyond the scope of the present work.\par
Finally, let us discuss the connection between this work and the concept of synthetic chiral light introduced in Ref. \cite{Ayuso:2019aa}. Our results show that a chiral molecule coupled to the photoionization continuum by a three-color field such that $\omega_1-\omega_2=\omega_3$ can show enantiosensitive EPs for appropriate field parameters. In particular, Eq. (5) shows that the enantiosensitivity is encoded in the three-photon matrix element $\Omega_{123}$, which takes opposite signs for opposite molecular enantiomers driven by the same field. A more transparent expression for the three-photon matrix element can be found by accounting for the random orientation of the molecule with respect to the laboratory frame in which we define the polarization of the laser field. After averaging over all orientations $\rho$, the three-photon matrix element is given by \cite{Barron:book}
\begin{equation}\langle \Omega_{123}\rangle_\rho=\left[\mathbf{d}_{1E_C}^*\cdot\left(\mathbf{d}_{2E_C}\times\mathbf{d}_{12}\right)\right]\left[\mathbf{F}^*_1\cdot\left(\mathbf{F}_2\times\mathbf{F}_3\right)\right].\end{equation}
We see that the three-photon matrix element factorizes into two pseudoscalars; the triple product of dipole matrix elements characterizes the handedness of the molecule, while the triple product of electric fields characterizes the handedness of the three-color field. In particular, the pseudoscalar of the field is nothing else but an example of the chiral correlation function $h^{(3)}$ defined in Refs. \cite{Ayuso:2019aa,Khokhlova:2022aa}, characterizing the handedness of a three-color field displaying chirality in the dipole approximation. That is, for an ensemble of randomly oriented chiral molecules, enantiosensitivity of EPs can be achieved only by using synthetic chiral light \cite{Ayuso:2019aa}. Moreover, factoring out the phases of each pseudoscalars we obtain
\begin{equation}\langle\Omega_{123}\rangle_\rho=|\chi_M||h^{(3)}|\cos(\phi_M-\phi_L)\end{equation}
where $\chi_M=\mathbf{d}_{1E_c}^*\cdot\left(\mathbf{d}_{2E_c}\times\mathbf{d}_{12}\right)$ is the molecular susceptibility \cite{Ayuso:2019aa} with phase $\phi_M=\phi_{2,E_c}+\phi_{1,2}-\phi_{1,E_c}$ (here $\phi_{ij}$ is the phase of the dipole matrix element $\mathbf{d}_{ij}$) and $\phi_L=\phi_2+\phi_3-\phi_1$ is the laser field phase, where swapping molecular enantiomer corresponds here to $\phi^R_M\rightarrow\phi^L_M+\pi$. We see therefore that in order to obtain enantiosensitivity of the EPs in the randomly oriented case we must have that $\phi_L\neq\phi_M+k\pi/2$, where k is an odd integer, and that enantiosensitivity is maximized when $\phi_L=\phi_M$. Obviously, a change in handedness of the field $\phi_L\rightarrow\phi_L+\pi$ will interchange the position of the EPs of the two enantiomers.\par
In conclusion, our work shows for the first time that it is possible to induce enantiosensitive EPs in a chiral non-Hermitian system. By considering a minimal three-level model, here meant to represent a chiral molecule coupled to the photoionization continuum by a three-color laser field, we have studied the position of the enantiosensitive EPs in parameter space, showing that by tuning the field parameters we can achieve optimal separation between EPs of opposite enantiomers. As an example on how to exploit such separation, we have shown that it is possible to induce the ASM effect for only one of the two enantiomers. Our results are independent of the initial conditions, as shown in the Supplementary Informations \cite{SuppNotes}, where we also study the position of the EPs and the ASM effect in the $(\Gamma_1,\Gamma_2)$ parameter space for fixed $\Delta$ and $\Omega_{12}$. Finally, we have shown the connection between the present work and the recently introduced concept of synthetic chiral light \cite{Ayuso:2019aa}. Note that in this work we have ignored the permament dipoles of the ground and excited state, which would result in additional on-diagonal terms in Eq. (1); yet, these can be easily included in our description, offering further control knobs for exploiting the physics here presented.\par
We stress that ASM is only ine of the remarkable effects associated to EPs. For example, the response of a system tuned to an EP to an external perturbation $\epsilon$ scales as $\epsilon^{1/2}$ (for a second order EP), in contrast to the $\epsilon$ scaling of Hermitian diabolical points \cite{Wiersig:2020aa}. One could therefore devise an EP-based chiral sensor that could detect with high-sensitivity molecules of a given handedness, even in a racemic solution. Finally, owing to the generality of the Hamiltonian in Eq. (1), we expect our work to be applicable to other systems such as waveguides, which could provide an ideal framework to study these effects.

\section*{Acknowledgments}

We gratefully acknowledge helpful discussions with Misha Ivanov, David
Ayuso, Margarita Khoklova and Emilio Pisanty. This
project has received funding from the EU Horizon 2020 (grant agree-
ment No 899794) and European Union (ERC, ULISSES, 101054696).

\bibliography{biblio}

\end{document}